\documentclass{article}

\usepackage{PRIMEarxiv}
\usepackage[utf8]{inputenc} 
\usepackage[T1]{fontenc}    
\usepackage{hyperref}       
\usepackage{url}            
\usepackage{booktabs}       
\usepackage{amsfonts}       
\usepackage{nicefrac}       
\usepackage{microtype}      
\usepackage{lipsum}
\usepackage{graphicx}
\graphicspath{{media/}}     
\usepackage{amsmath}

\begin{document}
\title{Practical study of optical stellar interferometry 
}

\author{
  P. Rodríguez-Ovalle, A. Mendi-Martos, A .Angulo-Manzanas, I. Reyes-Rodríguez, M.Pérez-Arrieta \\
  Aula Espazio-Gela, Escuela de Ingeniería de Bilbao \\
  UPV/EHU \\
  Plaza Ingeniero Torres Quevedo, 1, Bilbao, 48013, Bizkaia, Spain\\
   \And
  M. A. Illarramendi, A. Sánchez-Lavega \\
  Departamento de Física Aplicada, Escuela de Ingeniería de Bilbaon \\
  UPV/EHU \\
  Plaza Ingeniero Torres Quevedo, 1, Bilbao, 48013, Bizkaia, Spain\\
  \texttt{ma.illarramendi@ehu.eus}
 \\
}
\textbf{Paper already accepted by the American Journal of Physics}

\maketitle

\begin{abstract}
We present a technique to observe stellar inteferograms, as well as the detailed analysis of those created by three bright stars: Betelgeuse, Rigel, and Sirius. It is shown that the atmospheric turbulence is responsible for the reduction of the long-exposure fringe visibility of the obtained interference patterns.  By using different baselines in our interferometer, we are able to distinguish the decay of the visibility with the baseline, observe how different parameters such as the diameter or the distribution of the holes in our interferometer affect the pattern, and measure the atmospheric turbulence. This experiment can be performed by postgraduate students to gain practical experience in optical interferometry in astronomy.
\end{abstract}


\section{Introduction}\label{sec1}

In optical astronomy, interferometric techniques have been used ever since their discovery over a hundred years ago, for they enable increased angular spatial resolution \cite{1Michelson1921}. Interferometric techniques are used in both ground-based and space-based observatories. Of particular interest is their use in the study of the properties of stars, for example to measure their sizes, to characterize multiple stellar systems, and to precisely determine their positions in the sky and their motions (astrometry) \cite{2Labeyrie1978,3Monnier2003,4Lawson2000,5glindemann2011,6ESO}. A list of optical and infrared astronomical interferometers can be found in ref. \cite{7list}.  The technique is not simple to work with and is limited by the sky's atmospheric conditions at the time of observation, called 'seeing conditions'.

Teaching astronomical interferometry to undergraduate or graduate physics students is therefore not an easy task. The topic is usually presented briefly in optics courses \cite{8Hecht1974,9Born_pablo,10Pedrotti2017} and, when dealing with astronomy, most textbooks mainly focus on the classic Michelson interferometer \cite{11fundamAstron}. Few practical activities are carried out for training in these techniques at university level and this is done mostly in relation to astronomy, astrophysics and space science courses. At laboratory level, Michelson-type radio interferometer \cite{12Koda16} or interferometric experiments \cite{15Carbonell18} have been introduced. In the Master in Space Science and Technology of the Basque Country University UPV/EHU \cite{16SanchezLavega2014}, we have incorporated practical work into the educational program with optical telescopes coupled to laser sources with polymeric optical fibers simulating stars \cite{13Illarramendi2014,14Arregui2017}.

In this paper we describe a new interferometry experiment. By coupling different plates using several diaphragms, with variable apertures and separations, at the entrance of a 28 cm telescope, we have built a simple interferometer and, with it, we have observed three bright stars (Betelgeuse, Rigel and Sirius). We have analyzed the stellar interferograms using optical interferometry theory. It is shown that, very much like what happens in research-grade telescopes or in stellar interferometers \cite{18Roddier1981a},  atmospheric turbulence causes a reduction in the long-exposure fringe visibility by a factor that depends on the atmospheric coherence length, or Fried parameter ($r_0$). By studying the decay of the visibility with the baseline, we have estimated the Fried parameter for each case. The star sizes could not be estimated due to the small values of the baselines provided by the experiment \cite{1Michelson1921,2Labeyrie1978,3Monnier2003,4Lawson2000,5glindemann2011}, but our work describes a simple technique to obtain interference patterns produced by bright stars, the analysis of which introduces important physical concepts, such as spatial and temporal coherence.

\section{Theoretical background}\label{sec2}

\begin{figure}[ht]
    \centering
    \includegraphics[scale=0.75]{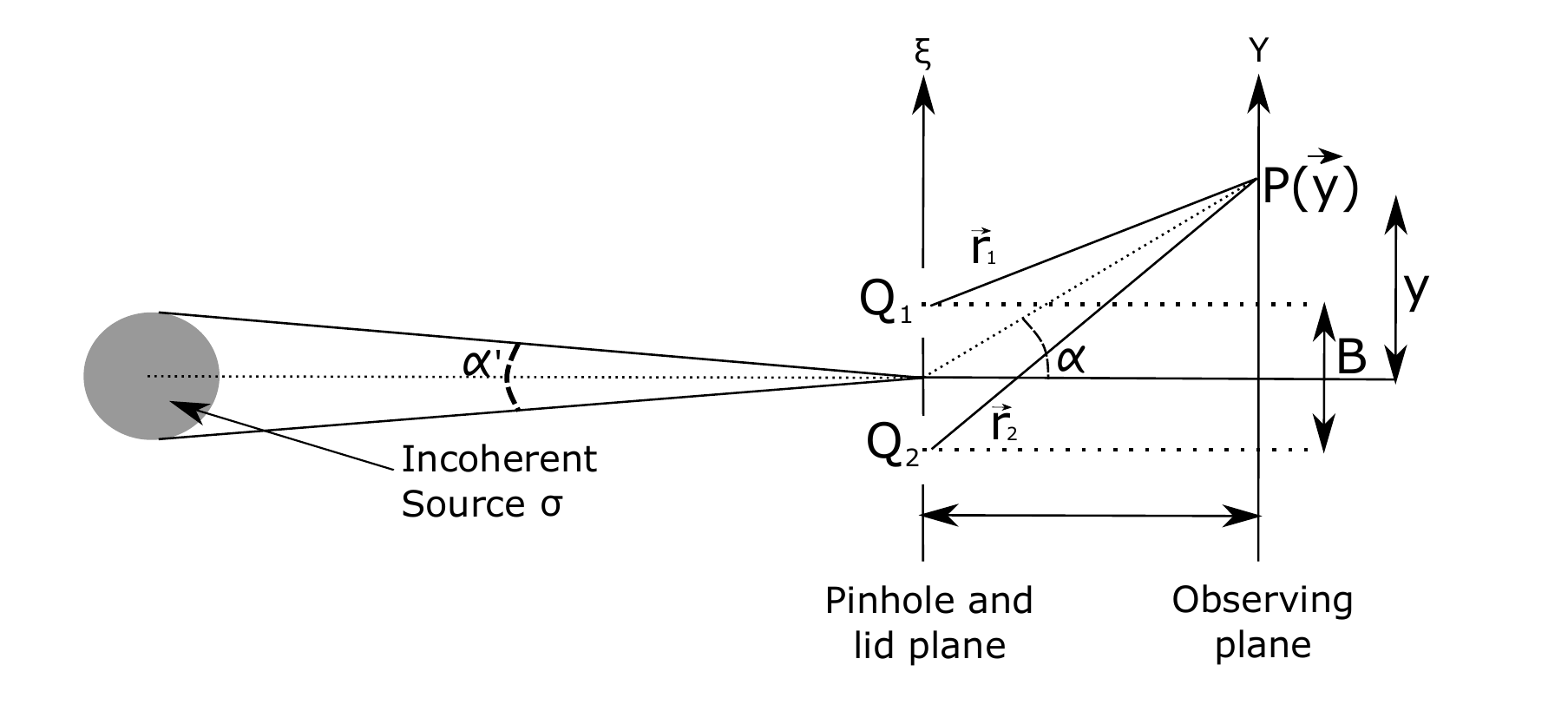}
    \caption{Simple scheme of a double pinhole stellar interferometer. The two pinholes ($Q_1$ and $Q_2$) of diameter $D$ are separated by a distance $B$, and are placed far away from the source.}
    \label{fig1}
\end{figure}

A simplified version of a stellar interferometer can be obtained by using a single telescope whose aperture is covered by a lid with two circular pinholes separated by a variable distance, called the baseline $B$. The working principle of this simple interferometer is the same as Young's double-slit experiment, where the beams emerging from each pinhole of diameter $D$ form interference fringes in the focal plane of the telescope or plane of observation. A schematic illustration of this interferometer is shown in Fig. \ref{fig1}. The quality of the interference fringes detected at the observation plane is measured by the fringe visibility or contrast $V$:

\begin{equation}
    V = \frac{I_{max}-I_{min}}{I_{max}+I_{min}} \text{    },
    \label{eq:eq1}
\end{equation}

\,

\noindent where $I_{max}$ and $I_{min}$ are the maximum and minimum intensities of the interference fringes, respectively. The visibility is scaled from 0 to 1, where 0 means no fringes and 1 denotes fringes with perfect contrast.  

Taking the simplest model to describe the light emission of a star, i.e. a circular, uniform, and spatially incoherent source emitting quasi-monochromatic light of wavelength $\lambda$, the diffraction-limited intensity at the plane of observation can be expressed as follows \cite{9Born_pablo}:

\begin{equation}
    I(\alpha)=I_{0}\left(\frac{J_{1}\left(\frac{\pi}{\lambda} \alpha D\right)}{\frac{\pi}{\lambda} \alpha D}\right)^{2}\left(1+V_{s} \cos \left(\frac{2 \pi}{\lambda} \alpha B\right)\right) ,
    \label{eq:eq2}
\end{equation}

\noindent with:

\begin{equation}
    V_{s}=2\left\lvert{\frac{J_{1}\left(\frac{\pi \alpha^{\prime} B}{\lambda}\right)}{\frac{\pi \alpha^{\prime} B}{\lambda}}}\right\rvert ,
\end{equation}

\noindent where $J_1$ is the first-order Bessel function of the first kind, $\alpha^{'}$ is the angular size of the source and $\alpha$ is the observation angle. The product $\alpha B$ is the optical-path difference for small values of $\alpha$, and $I_0$ is a constant related to the incident intensity. $V_s$ is the spatial fringe visibility for this simple model. $V_s$ does not depend on $\alpha$, and is inversely proportional to the source angular size and to the baseline distance. In fact, the steady decrease of $V_s$ from a value of 1 when  $(\pi \alpha^{'} B)/\lambda=0$ to a value of 0 when $(\pi \alpha^{'} B)/\lambda=1.22\pi$ allows the determination of the source size if $V_s$  is measured as a function of the baseline distance $B$. The easiest procedure to estimate the source diameter is to determine the lowest value of B for which the interference fringes disappear.  The reduction of $V_s$  as the source size or the baseline increases is a result of the spatial coherence of the light.  The function $\frac{J_1(\frac{\pi \alpha D}{\lambda})}{\frac{\pi \alpha D}{\lambda}}$ represents the irradiance distribution of the diffraction-limited response to a point source illuminating one circular pinhole, which is the Airy pattern. The total number of visible fringes is limited by the diffraction effect, as well as by the value of the visibility function $V_s$. 
If the baselines used in the measurements are short enough to provide very small values of $(\pi \alpha^{'} B)/\lambda$ , the fringe visibility is very close to 1 and, therefore, Eq. \ref{eq:eq2} could be simplified to:

\begin{equation}
    I(\alpha)=I_{0}\left(\frac{J_{1}\left(\frac{\pi}{\lambda} \alpha D\right)}{\frac{\pi}{\lambda} \alpha D}\right)^{2}\left(1+\cos \left(\frac{2 \pi}{\lambda} \alpha B\right)\right) ,
    \label{eq:eq3}
\end{equation}

\,

\noindent which is the diffraction-limited irradiance distribution produced by a point source emitting a  quasi-monochromatic light. If a finite spectral bandwidth $\Delta\lambda$ is taken into account for the source, the resulting fringe pattern can be inferred by adding up the interference patterns given by Eq. \ref{eq:eq3} at all wavelengths. This effect reduces the fringe visibility as the observation angle $\alpha$ or, equivalently, the time delay between interfering beams $\tau$ increases. The time delay for the interference fringes to vanish is called the coherence time $\tau_c$, which is defined as $\tau_c$=$\lambda^2$/c$\Delta\lambda$. Taking this definition into account, an approximate diffraction-limited irradiance distribution for a polychromatic light at the plane of observation can be written as:

\begin{equation}
    I(\alpha)=I_{0}\left(\frac{J_{1}\left(\frac{\pi}{\lambda} \alpha D\right)}{\frac{\pi}{\lambda} \alpha D}\right)^{2}\left(1+V_{t}\cos \left(\frac{2 \pi}{\lambda} \alpha B\right)\right) ,
    \label{eq:eq4}
\end{equation}

\noindent with:

\begin{equation}
    V_{t}= 1 - \alpha B \frac{\Delta \lambda}{\lambda^2} = 1 - m\frac{\Delta \lambda}{\lambda}
    \label{eq:eq4x}
\end{equation}

\noindent The exact expression for the temporal fringe visibility $V_t$ depends on the form of the spectral bandwidth \cite{9Born_pablo,10Pedrotti2017}. The parameter $m$ in Eq.\ref{eq:eq4x} is an integer number that indicates the order of interference. One of the consequences of observing a source with a significant bandwidth is the dependence of $V_t$ on the position on the observation plane.  In particular, its value decreases down to 0 as $\alpha$ increases up to $\alpha_e =\lambda^2 / B\Delta \lambda$, or as the interference order becomes $m = \lambda/\Delta\lambda$. This decrease in the visibility is therefore more pronounced when $\Delta \lambda$ is larger. The behavior of $V_t$ as a function of the wavelength distribution is a result of the temporal coherence of the light. If the values of $B$ and $\Delta \lambda$ used in the measurements result in very high values of $\alpha_e$, $V_t$ at the observation positions close to the optical axis would be nearly 1. In that case, Eq. \ref{eq:eq4} could also be simplified to Eq. \ref{eq:eq3}.

The preceding equations are only valid for an atmosphere without turbulence.  Turbulence causes the fringes to undergo random changes due to inhomogeneities encountered by light along its path to the interferometer. For long-exposure times and under the assumption $V_s$=$V_t$=1, the diffraction-limited irradiance distribution at the plane of observation can be expressed as follows:

\begin{equation}
    I(\alpha)=I_{0}\left(\frac{J_{1}\left(\frac{\pi}{\lambda} \alpha D\right)}{\frac{\pi}{\lambda} \alpha D}\right)^{2}\bigg(1+V_{a}\cos \left(\frac{2 \pi}{\lambda} \alpha B\right)\bigg) ,
    \label{eq:eq5}
\end{equation}

\noindent with: 

\begin{equation}
    V_{a}= exp\left[-3.44\left(\frac{B}{r_0}\right)^{5/3}\right]
    \label{eq:eq5x}
\end{equation}

\noindent  $V_a$ is the atmospheric fringe visibility of the detector's time-averaged interference pattern produced by the light that has passed through the Earth’s turbulent atmosphere \cite{17Fried65}. The Fried parameter, $r_0$, is an atmospheric coherence length that indicates the length over which the wavefront remains unperturbed and it is a measure of the seeing quality, i.e., the quality of the atmospheric conditions at the time of observation \cite{18Roddier1981a}. The smaller $r_0$ is, the larger the effects of turbulence on the propagating wave. $r_0$ varies as $\lambda^{6/5}$, so that it becomes smaller at shorter wavelengths which implies a more severe turbulence effect on the wavefront. Typical values for $r_0$ under good seeing conditions are 15-20 cm for visible wavelengths \cite{Hardy1998Jul}. In our case, the turbulence-induced random phase fluctuations of the fields drive visibility rapidly to zero as the baseline $B$ is increased (see expression of $V_a$ in Eq.\ref{eq:eq5x}); that is, atmospheric turbulence causes the light field to become spatially incoherent for baselines $B$ $\geq$ $r_0$. By analyzing the dependence of $V_a$ on baseline, we can estimate $r_0$, which allows us to characterize the atmospheric turbulence \cite{17Fried65}. 

\section{Experimental set-up}\label{sec3}

Interference fringes were obtained for the bright stars Betelgeuse, Rigel and Sirius. Betelgeuse has the largest angular size of any star in our night sky. It was the first star that was resolved by Michelson and Pease using stellar interferometry \cite{1Michelson1921}. As a red giant, its large size comes along with a relatively low temperature. Rigel, a blue super-giant star ($\beta$ Orion) has a surface temperature higher than 10000 K. Despite having an angular size much smaller than Betelgeuse’s, its high temperature makes it a bright object in the sky. Sirius has a much smaller radius compared with Betelgeuse or Rigel. As an A-type star, its temperature is close to 10000 K. However, this star is closer to Earth than the other two, at only 8.6 light years. Table \ref{tab1:star_properties} shows the stars' relevant properties. Sirius is a binary star, consisting of Sirius A (the brighter star) and Sirius  B. Since the ratio between the emitted brightness of both stars, known as the contrast factor $f$ of the binary star, is very small, Sirius can be approximated as a single star for this study.

\begin{table}[ht]
\begin{center}
\begin{minipage}{400pt}
\caption{Properties of the studied stars: their radius (in units of the Sun's radius $R_{\odot}$), their distance from Earth (in light-years), their angular aperture (mas = miliarcseconds), $f$ represents the brightness ratio between the two stars in the case of double stars, $m_v$ is their apparent magnitude (measure of the brightness), $T$ their surface temperature.}
\label{tab1:star_properties}
\begin{tabular}{lccccccc}
\hline\hline
\textbf{Star} & \multicolumn{1}{l}{\textbf{Radius ($\boldsymbol{\rm{R_\odot}}$)}} & \multicolumn{1}{l}{\textbf{Distance ($\boldsymbol{\rm{Ly}}$)}} & \multicolumn{1}{l}{\textbf{Size ($\rm{\textbf{mas}}$)}} & \textbf{f} & \multicolumn{1}{l}{\textbf{$\text{\textbf{m}}_\text{\textbf{V}}$}} & \multicolumn{1}{l}{\textbf{T($\rm{\textbf{K}}$)}} & 
\\
\hline
\textbf{Betelgeuse} & 887 & 497.95 & 54.04 & - & 0.52 & 3600\\
\textbf{Rigel} & 79 & 863 & 2.6 & - & 0.15 & 12100\\
\textbf{Sirius A} & 2 & 8.6 & 6.04 & 0.0001 & -1.46 & 9940\\
\hline\hline
\end{tabular}
\end{minipage}
\end{center}
\end{table}

\begin{figure}[ht]
    \centering
    \includegraphics[scale=0.75]{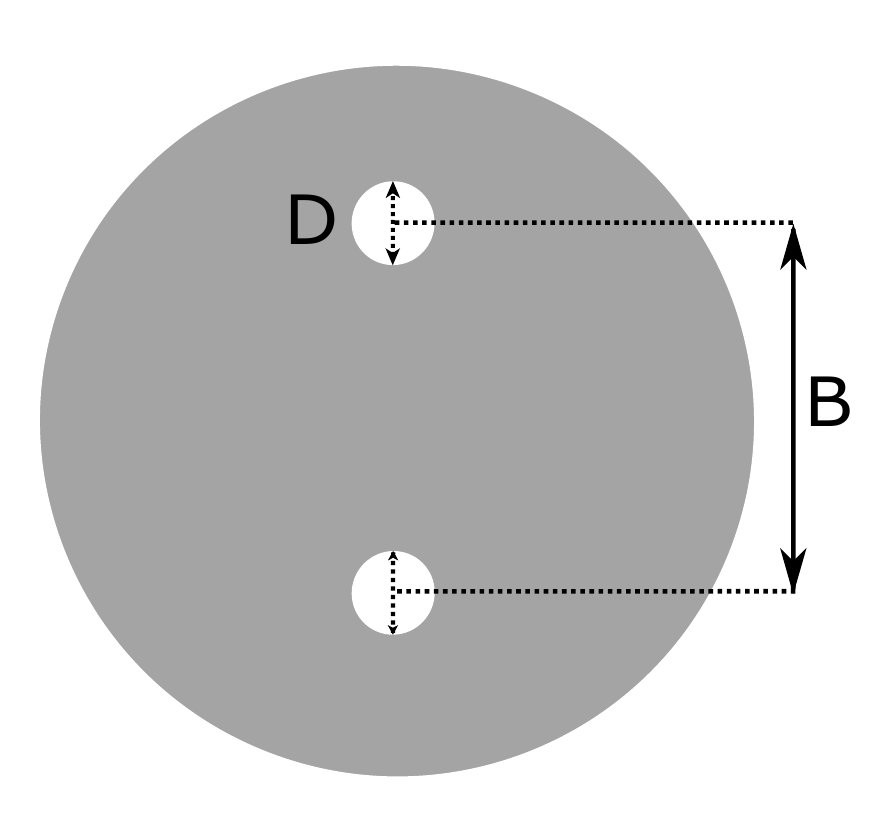}
    \caption{Diagram of the lid with two holes. The diameters $D$ of all the holes are equal for a given lid.}
    \label{fig2}
\end{figure}

The observations were carried out with a Celestron C11 XLT telescope with a 280 mm aperture and focal distance of 2800 mm from the Aula Espazio Gela Observatory in Bilbao (Spain) \cite{16SanchezLavega2014}. Once the telescope was well focused, its aperture was blocked by a lid with two or four circular pinholes. Only the case of two pinholes will be discussed in the main text. The restults for the four pinholes set-up are given in the Supplementary Material (See online supplemental material at \textbf{url to be inserted by AIPP}). Fig. \ref{fig2} illustrates the arrangement of the pinholes on the lids. Several two-pinhole lids were made with different combinations of pinhole diameters $D$ and baselines $B$. The lids were made of a robust cardboard and the pinholes were carefully shaped in order to minimize errors. All the chosen values for the pinhole diameters $D$ allowed us to observe the interference fringes clearly. Table \ref{tab2:Exp_param} summarizes the lid parameters used for the three stars.  

\begin{table}
\begin{center}
\begin{minipage}{400pt}
\caption{Values of the experimental parameters used for each observation night. The absolute error for the values of D and B is 1 mm. For each filter, the bandwidth is indicated.}\label{tab2:Exp_param}

\resizebox{0.495\textwidth}{!}{
\begin{tabular}[t]{cccc}
\hline\hline
\multicolumn{4}{c}{\textbf{BETELGEUSE}} \\ \hline
\textbf{D (mm)} & \textbf{B (mm)} & \textbf{Filter} & \textbf{Date (yy-mm-dd)} \\ \hline
30 & 133 & H$\alpha$ 35 nm & 17-03-15 \\
30 & 238 & H$\alpha$ 35 nm & 17-03-15 \\
51 & 147 & H$\alpha$ 35 nm & 17-03-15 \\ \hline

\hline \hline
\multicolumn{4}{c}{\textbf{RIGEL}} \\ \hline
\multicolumn{1}{l}{\textbf{D (mm)}} & \multicolumn{1}{l}{\textbf{B (mm)}} & \multicolumn{1}{l}{\textbf{Filter}} & \multicolumn{1}{l}{\textbf{Date (yy-mm-dd)}} \\ \hline
51 & 147 & H$\alpha$ 35 nm & 17-03-15 \\
51 & 222 & H$\alpha$ 35 nm & 17-03-15 \\
\hline\hline
\end{tabular}
}
\resizebox{0.495\textwidth}{!}{
\begin{tabular}[t]{cccc}
\hline \hline
\multicolumn{4}{c}{\textbf{SIRIUS}} \\ \hline
\multicolumn{1}{l}{\textbf{D (mm)}} & \multicolumn{1}{l}{\textbf{B (mm)}} & \multicolumn{1}{l}{\textbf{Filter}} & \multicolumn{1}{l}{\textbf{Date (yy-mm-dd)}} \\ \hline
22 & 220 & H$\alpha$ 7 nm & 17-03-10 \\
22 & 158 & H$\alpha$ 7 nm & 17-03-10 \\
61 & 178 & H$\alpha$ 7 nm & 17-03-10 \\
30 & 238 & H$\alpha$ 35 nm & 17-03-23 \\
30 & 133 & H$\alpha$ 35 nm & 17-03-23 \\
50 & 222 & H$\alpha$ 35 nm & 17-03-23 \\
65 & 162 & H$\alpha$ 35 nm & 17-03-23 \\
66 & 162 & H$\alpha$ 35 nm & 17-03-29 \\
65 & 205 & H$\alpha$ 35 nm & 17-03-29 \\
90 & 187 & H$\alpha$ 35 nm & 17-03-29 \\
66 & 162 & H$\beta$ 8.5 nm & 17-03-29 \\
90 & 187 & H$\beta$ 8.5 nm & 17-03-29 \\ 
\hline\hline
\end{tabular}
}
\end{minipage}
\end{center}
\end{table}

Images of the fringe pattern were acquired with a camera attached to the focal plane of the telescope. We used a DMK41AU02 canera equipped with the Sony ICX205AL CCD chipset with a pixel size of 4.65 $\mu m$. To improve the quality of the images, we processed them by stacking video sequences for every observation with the same methodology described in previous papers \cite{19Rojas2017,20SanchezLavega2019}. Hence, we can obtain images with a better quality and remove the random noise that could appear in individual frames due to the effects of atmospheric turbulence and telescope vibrations. To obtain resolved fringes at different wavelengths, we used two $H_{\alpha}$ ﬁlters centered on a wavelength of 656.3 nm with bandwidths of 7 and 35 nm respectively, and a $H_{\beta}$ filter with a 8.5 nm bandwidth centered at 486.1 nm \cite{21baader7nm}. 

We used very long exposure times, ranging from 0.3 s to 2.4 s so that suitable fringe images were detected. This fact adversely affected the measurements due to the additional effect of different turbulence scales in the atmosphere, which usually change on timescales longer than a few miliseconds. The obtained long-exposure images were converted to digital values, according to the camera range. The brightness levels from the interference patterns were digitized into 512 levels. Those levels were treated with the free software ImageJ, which allowed us to obtain the profile of the interference patterns \cite{22ImageJ}. The software also allowed us to determine the values of the irradiance $I_{max}$ and $I_{min}$ to calculate the visibility of the fringe pattern. In order to show the photometric cuts as a function of the angular size in the focal plane, the distances in pixels were transformed to radians as seen through the 2800 mm focal length of our telescope, by taking into account the magnifying effect of the telescope's Barlow lens.

\section{Results and discussion}\label{sec4}

\begin{figure}  
    \centering
    \includegraphics[scale=0.59]{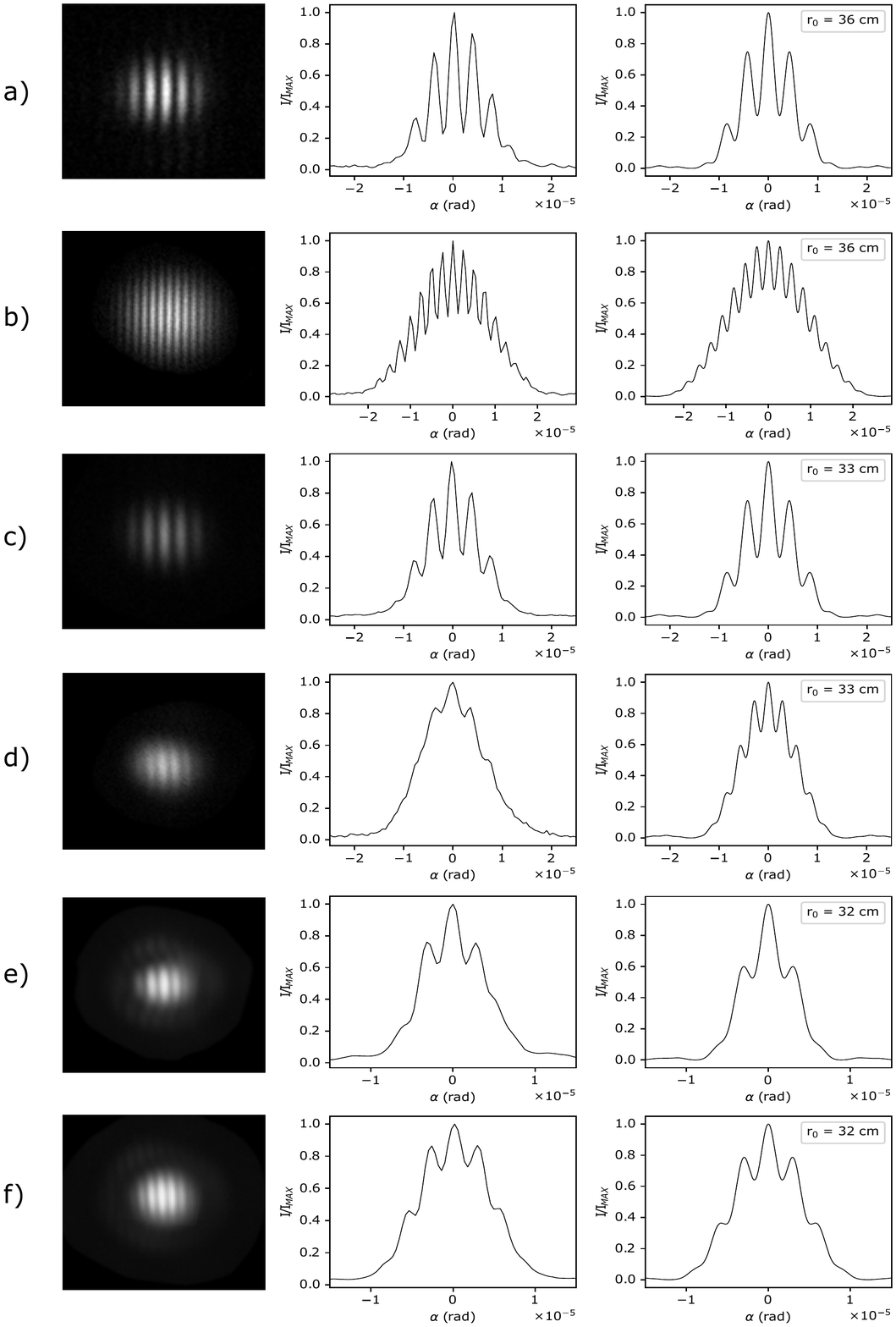}
    \caption{Two-pinhole interference pattern produced with the $H_{\alpha}$ filter, $\Delta \lambda$ = 35nm. On the left and in the center, the interference patterns obtained experimentally and the corresponding photometric cuts; on the right, the theoretical curves calculated from Eq. \ref{eq:eq5} with $\lambda$ = 656.3 nm, and the corresponding values of $D$, $B$, and $r_0$. (a) Betelgeuse star, $D$ = 51 mm, $B$ = 147 mm; (b) Betelgeuse star, $D$ = 30 mm, $B$ = 238 mm; (c) Rigel star, $D$ = 51 mm, $B$ = 147 mm; (d) Rigel star, $D$ = 51 mm, $B$ = 222 mm; (e) Sirius star, $D$ = 90 mm, $B$ = 187 mm; (f) Sirius star, $D$ = 66 mm, $B$ = 205 mm.}
    \label{fig3}
\end{figure}

\begin{figure}[ht]
    \centering
    \includegraphics[scale=0.65]{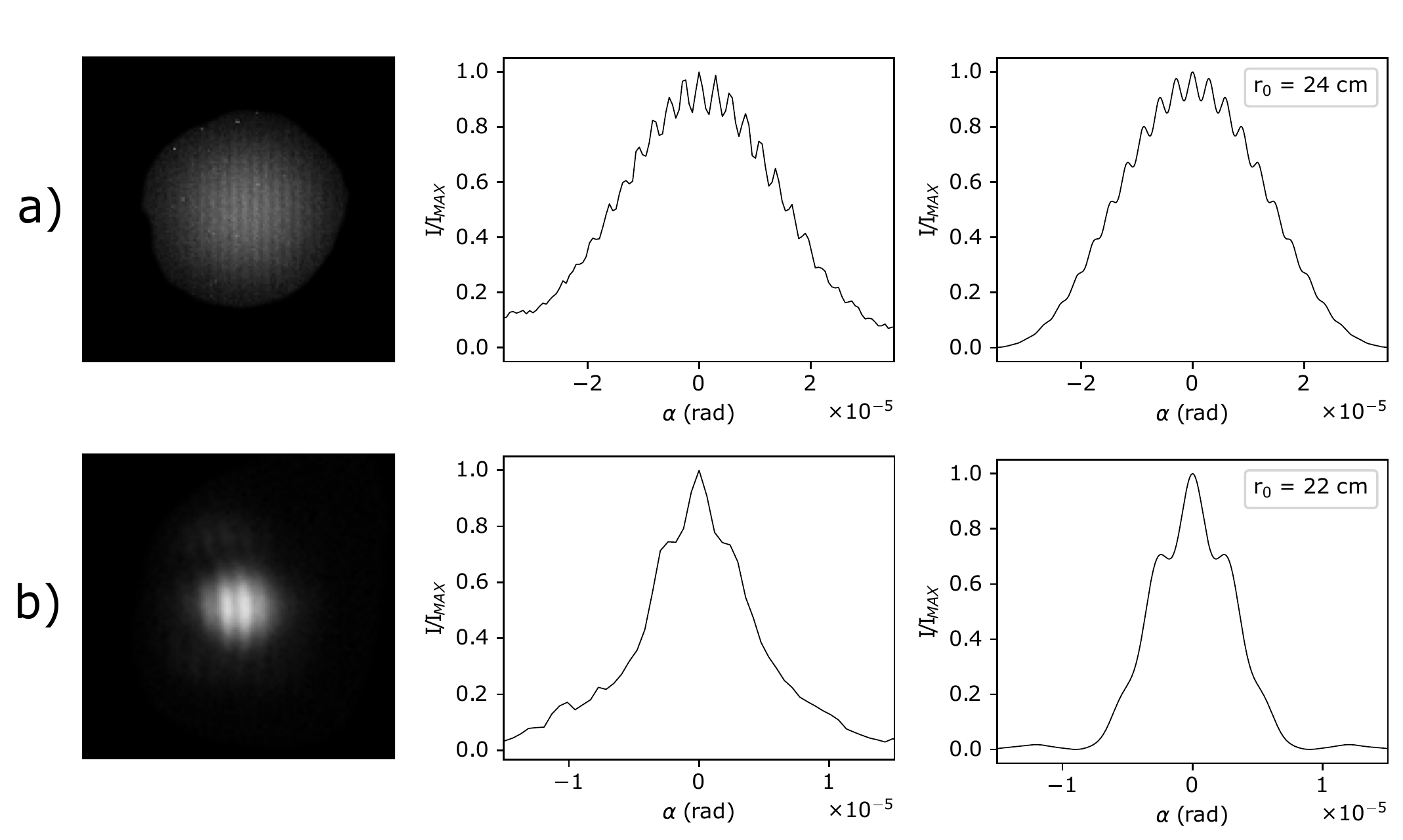}
    \caption{Two-pinhole interference patterns produced by Sirius with the narrow $H_{\alpha}$ filter ($\lambda$ = 656.3 nm, $\Delta\lambda$ = 7 nm, $D$=22 mm, $B$=220 mm, top), and the $H_{\beta}$ filter ($\lambda$ = 486.1 nm, $\Delta\lambda$ = 8.5 nm, $D$=66 mm, $B$=162 mm, bottom). On the left and in the center, the interference patterns
    obtained experimentally and the corresponding photometric cuts. On the right the theoretical curve calculated
    from Eq. \ref{eq:eq5}.
}
    \label{fig4}
\end{figure}

In this Section, we will study the interference patterns obtained for the different values of baselines ($B$) and spectral bandwidths ($\Delta \lambda$) displayed in Table \ref{tab2:Exp_param}. Using such small values of $B$ and such narrowband filters leads to the atmospheric turbulence being the main cause of the reduction in the visibility of the observed fringe patterns. Fig. \ref{fig3} shows examples of such interference patterns and the corresponding relative irradiance profiles obtained for different combinations of $D$ and $B$ with the 35 nm bandwidth-$H_{\alpha}$ filter and the two pinholes lid. Figs. \ref{fig3}(a) and (b) correspond to the Betelgeuse star, Figs. \ref{fig3}(c) and (d) to Rigel, and Figs. \ref{fig3}(e) and (f) to Sirius. As can be seen, the fringe patterns obtained with the double-pinhole lid deviate little from the diffraction-limited interference patterns. We can notice both that the interference fringes are modulated by the larger Airy pattern generated by the circular apertures and that the spatial frequency of the fringes increases with $B$. We also notice that the fringe visibility is poor in all cases ($V$ $<$ 0.88). Fig. \ref{fig4} displays the interference fringes from Sirius obtained with the narrowest $H_{\alpha}$ filter and the $H_{\beta}$ filter, which also have a low-visibility. The interference patterns obtained with other experimental parameters (see Table \ref{tab2:Exp_param}) are very similar to those shown in Figs. \ref{fig3} and \ref{fig4}. Since the conditions to approximate $V_s$ and $V_t$ to 1 are satisfied in our experiment, the obtained visibilities are mainly due to the effect of the atmospheric turbulence. On the one hand, the very short baselines are such that $\frac{\pi\alpha^{'}}{\lambda}$ $<$ 0.4 rad, thus providing that $V_s$ is very close to 1. Note that the fact that the visibility does not decrease with B implies that we cannot estimate the source size. For Betelgeuse, which is the largest star analyzed in this work, interference fringes would disappear entirely with a baseline of about 3 meters \cite{1Michelson1921}, which is unrealistic in most practical setups.  On the other hand, the order of interference for which the fringe visibility goes to zero using the broadest filter ($H_{\alpha}$ filter with $\Delta \lambda$ = 35 nm) is 18, which is far away from the maximum of interference. This implies that $V_t$ at the observation positions near the optical axis can also be approximated by 1. Therefore, we can conclude that the cause of the reduction in the visibility of our patterns is due to atmospheric turbulence and use Eq. \ref{eq:eq5} to describe the irradiance distribution of the interference fringes. From the analysis of the dependence of the visibility values on the baseline, we will be able to derive the Fried parameter for each of our experimental conditions.

\begin{figure}[ht]
    \centering
    \includegraphics[scale=0.75]{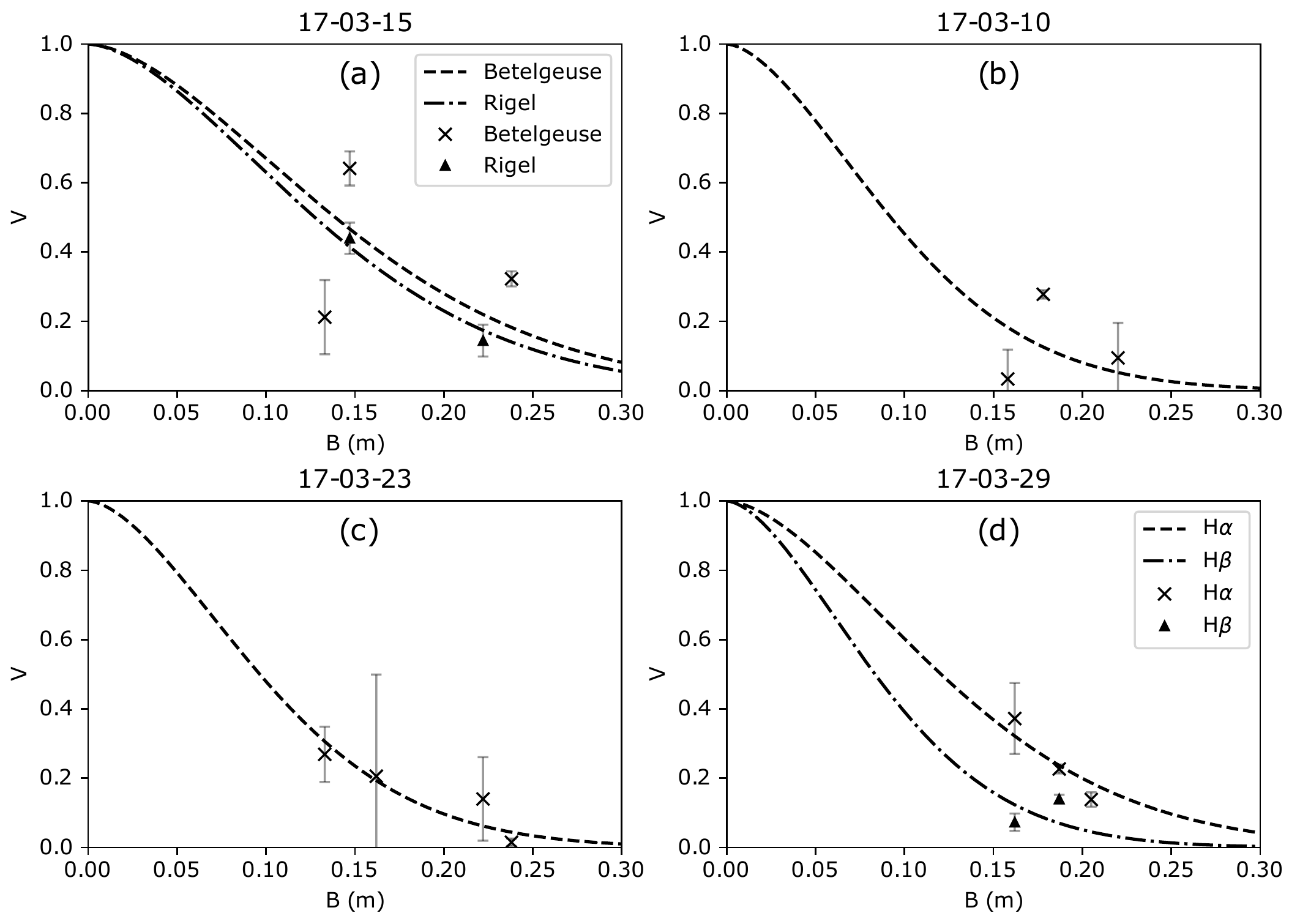}
    \caption{Visibility measurements as a function of the two-pinhole baseline distance obtained for the three stars. The dashed lines are the fittings of the data to the expression of  $V_a$.  Data obtained from (a) Betelgeuse and Rigel on March 15, 2017; (b) Sirius on March 10, 2017; (c) Sirius on March 23, 2017; (d) Sirius with $H_{\alpha}$ and $H_{\beta}$  on March 29, 2017.}
    \label{fig5}
\end{figure}

Fig. \ref{fig5} shows the visibility $V$ of the stars analyzed as a function of the two-pinhole distance. The data plotted in each figure have been obtained on the same night. The values of the fringe visibility are quite low and they tend to diminish as the baseline increases, as expected from Eq. \ref{eq:eq5x}. In the calculation of $V$, $I_{max}$ has been determined from the irradiance of the central maximum and $I_{min}$ from the average of the two adjacent minima. All irradiance measurements have been corrected for the background irradiance and the errors in the visibility values were estimated from the standard deviations calculated from five independent measurements taken for each interference patterns. It can be shown that the error in the visibility arising from the uncertainties of the baselines and pinhole diameters ($\pm$ $1$ $mm$) is negligible in comparison with the standard-deviation error.


By fitting the expression of $V_a$ in Eq. \ref{eq:eq5x} to the measured visibility data plotted in Fig. \ref{fig5}, we have estimated the Fried parameter $r_0$ for each observation night at the wavelengths used (Table \ref{tab3:rzero_results}), i.e. at 656.3 nm for the fringes obtained with the $H_{\alpha}$ filters and at 486.1 nm for those captured with the $H_{\beta}$. The highest values of the Fried parameter were obtained for the night of March 15, 2017. It can be noticed that the $r_0$ values at 656.3 nm obtained from the interference fringes of Sirius are lower than those obtained from Betelgeuse and Rigel. This could be due to stronger atmospheric turbulence on the corresponding observation nights. The longer exposure times used to obtain the interference fringes from Sirius due to its lower emission in the spectral range of the $H_{\alpha}$ filter, could also affect the values obtained. On the other hand, the spectral dependence of the two $r_0$ values calculated from Fig. \ref{fig5} (d), at 656.3 nm and 486.1 nm agrees very well with the theoretical spectral dependence of the Fried parameter, that is, $r_0 \propto \lambda^{6/5}$. By using the proportionality constant ($0.0133 \pm 0.0008\;cm/nm^{6/5}$) calculated from the value of $r_0$ at $\lambda$ = 656.3 nm, we get $r_0$ = 22 $\pm$ 2 cm at $\lambda$ = 486.1 nm, which is consitent with the experimental value displayed in Table \ref{tab3:rzero_results}.  Using the same proportionality constant, we estimate that $r_0$ at 500 nm is 23 $\pm$ 2 cm. This value is slightly higher than the typical Fried parameter at a good observation site \cite{Hardy1998Jul}. This result, together with the larger values obtained for $r_0$ on the other observation nights, indicates that the atmospheric effect on the experimental images was small in all cases. In all of them, the obtained values for $r_0$ are of the order of the size of the telescope and, of course, larger than the diameter of the holes. Hence, these values of the atmospheric coherence lengths $r_0$, are in agreement with the fact that the shape of the interference fringes are dominated by the diffraction effect, as was shown in Figs. \ref{fig3} and \ref{fig4}. Nevertheless, more accurate results for Fried parameters could be obtained with a systematic study of the dependence of the visibility with $B$.

\begin{table}[t!]
\centering
\begin{minipage}{400pt}
\caption{Fried parameters at the corresponding wavelength for each observation night obtained from the fittings of the experimental points of Fig \ref{fig5}. Errors for the Fried parameters have been estimated from the fittings using Eq. \ref{eq:eq5}.}
\label{tab3:rzero_results}
\begin{tabular}{ccccc}
\hline\hline
\textbf{Date (yy-mm-dd)} & \textbf{Star}          & \textbf{$r_0$ (cm)} & \textbf{$\lambda$ (nm)} & \textbf{$\Delta\lambda$ (nm)}\\ \hline
17-03-15      & Betelgeuse             & \;\,36\;$\pm$\;10   &  656.3 & 35 \\
17-03-15      & Rigel                  & 33\;$\pm$\;2   &  656.3 & 35 \\
17-03-10      & Sirius                 & 24\;$\pm$\;5   &  656.3 & 7  \\
17-03-23      & Sirius                 & 25\;$\pm$\;2   &  656.3 & 35 \\
17-03-29      & Sirius                 & 32\;$\pm$\;2   &  656.3 & 35\\
17-03-29      & Sirius                 & 22\;$\pm$\;4   &  486.1 & 8.5\\          \hline\hline      
\end{tabular}
\end{minipage}
\end{table}

In the third column of Fig. \ref{fig3} and Fig. \ref{fig4}, we have included the theoretical photometric curves obtained from Eq. \ref{eq:eq5}. As can be seen, the experimental and theoretical photometric curves are in good agreement regarding the variation of the number of visible fringes with different combinations of $D$ and $B$. The fringe visibilities are also well described by Eq. \ref{eq:eq5x}. If the fringe visibility is high enough, the number of visible fringes can be easily calculated from the angular radius of the Airy disk ($1.22\lambda/D$) and the angular fringe spacing ($\lambda/B$), in the same way as in the diffraction–limited interference fringes. For instance, one can predict that the change in $D$ and $B$ for the two fringe patterns produced by Betelgeuse (Figs. \ref{fig3}(a) and (b)) results in a strong increase in the total number of visible fringes from 7 to 19, which is clearly seen in the experimental curves. In contrast, if the fringe visibility is low, it is more difficult to analyze the effect of changing the values of $D$ and $B$ on the interference patterns, as in the case of the Sirius interferograms (Figs. \ref{fig3}(e) and (f)). Likewise, if the fringes were only diffraction-limited, we should observe 25 fringes in the interference pattern displayed in Fig. \ref{fig4} (a) and 5 fringes in that displayed in Fig. \ref{fig4} (b).  In spite of the low visibility values of the two interference patterns displayed in Fig. \ref{fig4}, the theoretical predictions indeed agree quite well with the experimental results.  

\,

\section{Conclusion}\label{sec5}

In this work, we show a simple experiment in which long-exposure interference patterns produced by three stars, Betelgeuse, Rigel and Sirius, can be detected using a bandpass filter and a digital camera coupled to a small telescope obscured by lids with two (or more) pinholes. It has been demonstrated that the obtained long-exposure fringe patterns produced by the stars can be very well described by diffraction-limited interference patterns produced by a quasi-monochromatic point source, but with a fringe visibility reduced due to the atmospheric turbulence. Through the analysis of the fringe visibility as a function of the two-pinhole baseline, we have been able to characterize the affectation of the atmospheric turbulence to astronomical observations by calculating the Fried parameter for each observation night. The experiment can be useful for students and/or teachers in universities. Its simplicity and the interest of the obtained results make this experiment ideal for postgraduate subjects, such as: Astrophysics, Astronomy, Optics or Astronomical and Space Signal Processing. In addition to showing the principle of operation of the Michelson stellar interferometer, the experiment underlines important concepts related to spatial interferometry, such as spatial and temporal coherence and astronomical seeing.


\section*{acknowledgments}\label{sec6}

 This work has been supported by a research grant from Diputación Foral de Bizkaia-Bizkaiko Foru Aldundia to Aula EspaZio Gela at UPV/EHU, with additional support in the frame of Spanish project PID2019-109467GB-I00 (MINECO/FEDER, UE), with additional support in the frame of the Grant PID2019-109467GB-I00 funded by MCIN/AEI/10.1303 /501100011033/ and by Grupos Gobierno Vasco IT1742-22. 


\section*{Competing interests}\label{sec7}

The authors declare no competing interests.

\section*{Availability of data and materials}\label{sec8}

The datasets used and/or analyzed during the current study available from the corresponding author upon reasonable request.


\appendix
\renewcommand\thefigure{\thesection.\arabic{figure}}    
\section{Suplementary materials: 4-pinhole study case}\label{sec9}
\setcounter{figure}{0}

Another type of interferometer can be made by covering the telescope with a lid having more than two pinholes. This arrangement could provide more baselines, with different modules and orientations, thus allowing to conduct additional measurements and therefore to obtain more information \cite{2Labeyrie1978,3Monnier2003,4Lawson2000,5glindemann2011}. In astronomical interferometers,  as many different baselines as possible are required in order to get a good quality image. 

\begin{figure}[htp]
    \centering 
    \includegraphics[scale=0.45]{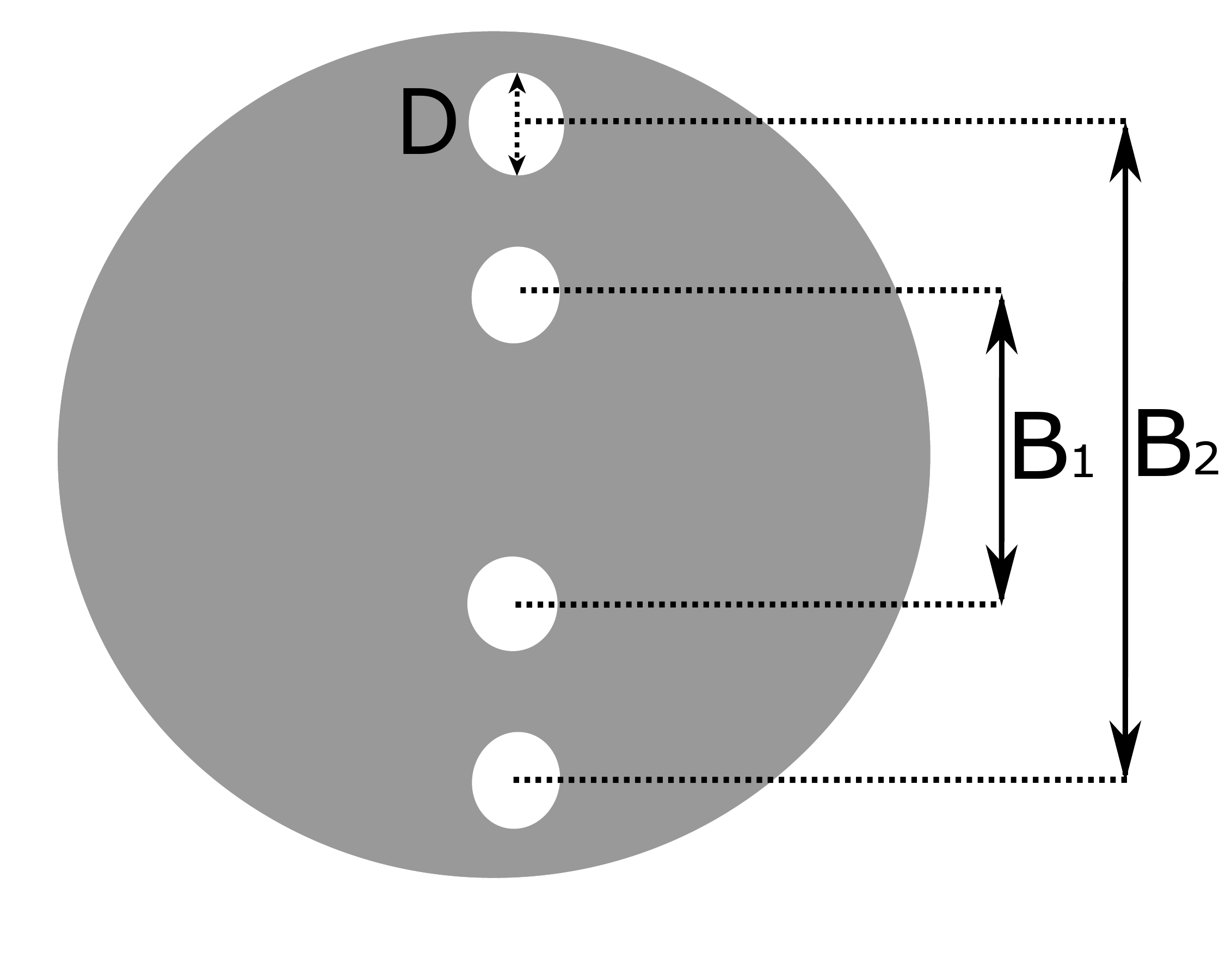}
    \caption{Diagram of the lid with four holes: $D$ = 30 mm, $B_1$ = 133 mm, $B_2$ = 238 mm. The diameters $D$ of all the holes are equal. }
    \label{figA1}
\end{figure}

The interference pattern produced by four pinholes can be generated by the superposition of all light beams coming from each hole. The general expression for the irradiance distribution can be given in terms of six different visibilities describing the correlation of the optical fields for each pair of holes. We have worked out the diffraction-limited interference pattern produced by four pinholes placed as shown in Fig. \ref{figA1}. Since two of the baselines are equal in this case, only four visibilities are necessary. Following Eq. 7, the intensity is given by:

\begin{equation}
\label{eq:eq6}
\begin{aligned}
    &I(\alpha)=I_{0}\left(\frac{J_{1}\left(\frac{\pi}{\lambda} \alpha D\right)}{\frac{\pi}{\lambda} \alpha D}\right)^{2}\bigg(4+4 V_{1} \cos \bigg(\frac{\pi}{\lambda} \alpha\left(B_{2}-B_{1}\right)\bigg) \,\, + \\
    &4 V_{2} \cos \bigg(\frac{\pi}{\lambda} \alpha\left(B_{2}+B_{1}\right)\bigg)
    +2 V_{3} \cos \left(\frac{2 \pi}{\lambda} \alpha B_{1}\right)+2 V_{4} \cos \left(\frac{2 \pi}{\lambda} \alpha B_{2}\right)\bigg) .
\end{aligned}    
\end{equation}

\,

\noindent For very short baseline distance ($B$ $<<$$\lambda$/$\pi$ $\alpha'$) and a quasi-monochromatic light, the visibilities can be determined by the Earth’s atmosphere turbulent effects by using the expression of $V_a$ shown in Eq. 8. In an atmosphere without turbulence, Eq. \ref{eq:eq6} could be simplified to: 

\begin{equation}
    I(\alpha)=I_{0}\left(\frac{J_{1}\left(\frac{\pi}{\lambda} \alpha D\right)}{\frac{\pi}{\lambda} \alpha D}\right)^{2}\bigg(\cos \left(\frac{\pi}{\lambda} \alpha B_{1}\right)+\cos \left(\frac{\pi}{\lambda} \alpha B_{2}\right)\bigg)^{2}
    \label{eq:eq7}
\end{equation}

\noindent If, in addition, the holes are equally spaced, $B_1=B_2/3=B$, by applying trigonometric relations, Eq. \ref{eq:eq7} could be simplified to \cite{8Hecht1974,9Born_pablo,10Pedrotti2017}:

\begin{equation}
    I(\alpha)=I_0\left(\frac{J_1\left(\frac{\pi}{\lambda} \alpha D\right)}{\frac{\pi}{\lambda} \alpha D}\right)^2 \frac{\sin ^2\left(4\left(\frac{\pi}{\lambda} \alpha \mathrm{B}\right)\right)}{\sin ^2\left(\frac{\pi}{\lambda} \alpha \mathrm{B}\right)} .
    \label{eq:eq8}
\end{equation}

\begin{figure}[ht!]
    \centering
    \includegraphics[scale=0.75]{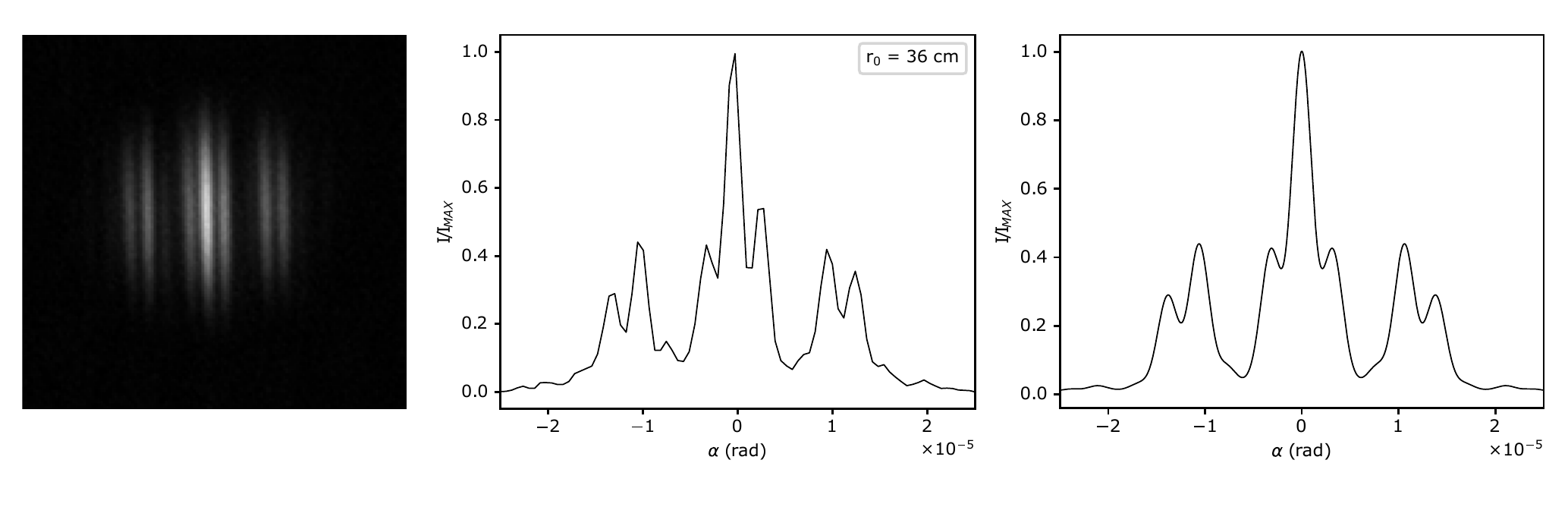}
    \caption{Four-pinhole interference pattern produced by Betelgeuse with the $H_{\alpha}$ filter of 35 nm (17-03-15). On the left and in the center, the interference pattern obtained experimentally and its photometric cut. On the right, the theoretical curve calculated from Eq. \ref{eq:eq6} with $\lambda$ = 656.3 nm, $D$= 30 mm, $B_1$ = 133 mm $B_2$= 238 mm, $V_1$ = 0.87, $V_2$ = 0.32, $V_3$ = 0.18, $V_4$ = 0.52.}
    \label{fig6_general_scheme}
\end{figure}

Fig. \ref{fig6_general_scheme} shows the interference pattern produced by Betelgeuse on the night of 17-03-15, using the lid with the four holes and the $H_{\alpha}$ filter of 35 nm in bandwidth. We can clearly appreciate the effect of having more than two holes on the interference pattern: the emergence of secondary maxima of irradiance and the narrowing and brightening of the fringes (principal maxima of irradiance) \cite{8Hecht1974,9Born_pablo,10Pedrotti2017}. Since the baselines are very short and the light can be assumed to be quasi-monochromatic, the obtained interference pattern can be described from Eq. \ref{eq:eq6} using the parameters of the measurement and the four visibilities calculated from the expression of $V_a$ of Eq. 8. The experimental photometric cut and the theoretical prediction have been plotted in the center and right panels of Fig. \ref{fig6_general_scheme}, respectively. The four visibilities have been calculated by using the equation of $V_a$ with the four different baseline distances of the lid and with the value of the Fried parameter estimated previously from Betelgeuse measurements carried out in the same night. As can be seen, a very good agreement is obtained between the theoretical and experimental curves, confirming the value of the Fried parameter.

\bibliography{thebibliography}

\end{document}